\documentstyle[11pt,twoside,paspconf,psfig]{article}

\begin{document}

\title{A Re-Examination of the Origins of the Peculiar Velocity Field
Within the Local Supercluster}

\author{David Burstein}
\affil{Arizona State University, Dept. of Physics and Astronomy,
Box 871504, Tempe, AZ 85287-1504}

\begin{abstract}

    The recent re-evaulation of the peculiar velocity field outside the Local
Supercluster (Dekel et al. 1999) has permitted a re-examination of the origins
of the peculiar velocity field within the Local Supercluster using the Mark
III Catalog of Galaxy Peculiar Velocities.  It is shown that the large-scale
coherent pattern of peculiar velocities within the LSC are well-fit by a
combination of the Outside-Region-(generated)-Motions (O-R-M) from the Potent
model with a Virgocentric infall pattern that produces 220~km~s$^{-1}$ of
Virgocentric infall at the Local Group towards the Virgo cluster moving at
88~km~s$^{-1}$ towards the Local Group.  The part of the LG CMB motion this
model cannot fit is that perpendicular to the Supergalactic plane. On what
size scale this SGZ motion of the Local Group is shared by neighboring
galaxies cannot be determined from the present data set, but can be found
if we can accurately measure galaxy distances close to the Galactic plane.

\end{abstract}

\section{The Four Issues to be Addressed}

    In the Mark III Catalog (Willick et al. 1995, 1996) the distance of the
Virgo cluster (VC) is $\rm D_{\rm VC} = 1330 \pm 52$~km~s$^{-1}$.  The
heliocentric radial velocity of the VC ($\rm V_{\rm VC}(helio)$) from Huchra 
(1996) is $1094 \pm 35$~km~s$^{-1}$.  Combine this with the heliocentric
peculiar motion of the Sun in the Cosmic Microwave Background (CMB) frame of
368.6 km s$^{-1}$ towards l = 264.7$^\circ$, b = 48.2$^\circ$, yields a
CMB radial velocity for the VC ($\rm V_{\rm VC}(CMB)$) of $1420 \pm
35$~km~s$^{-1}$. If we combine $\rm V_{\rm VC}(CMB)$ with the peculiar
velocity of the Local Group (LG) in the CMB frame (372~km~s$^{-1}$) in the
direction of the VC, we find the LG and the VC move towards each other in
the CMB reference frame with a mutual velocity of $\rm V_{\rm pec}(LG:VC) = 
282 \pm 63$~km~s$^{-1}$.  Four issues can be addressed from these data:

     1) How much of the CMB motion between the LG and the VC is
generated by the VC itself (i.e., classic ``Virgocentric infall'' motion),
how much is generated by masses on larger size scales?

     2) The peculiar velocities of the galaxy groups defined to be 
within the Coma-Sculptor (C-S) Cloud by Tully (1987) can be demonstrably shown 
to share most of the LG CMB peculiar velocity (e.g. Faber \& Burstein
1988; Tully 1988).  Much of this motion ($\sim 300$~km~s$^{-1}$) is directed
perpendicular to the Supergalactic plane.  Why?  How?

     3) As viewed in the LG reference frame, galaxy groups within the C-S Cloud
are moving towards each other.  How much of these motions are generated by 
masses within the C-S Cloud and how much are externally generated?

     4) In the direction towards the Ursa Major (Umaj) cluster (+X, +Y SG
coordinates), all galaxies with $\rm D \le 1800$~km~s$^{-1}$ have 
peculiar velocities moving towards the LG.  How much of these inward motions
are generated by mass within the Local Supercluster (LSC), and how much by mass
outside it?

     Why rexamine these issues now?  Based on the Mark III database and on a
new POTENT methodology, Dekel et al. (1999) have generated a more accurate
model of the peculiar motions generated {\it within} the LSC by masses {\it
outside} the LSC.  For the present paper, Ami Eldar and Avishai Dekel have
kindly calculated from this model the ``outside--region--(generated)--motions''
(O-R-M) for galaxies within a distance of 3000~km~s$^{-1}$ from the masses
outside this region.  For this analysis, we use only the spiral galaxies in
the Mark III database.  We use the CMB velocity reference frame only for
galaxies outside the C-S Cloud.  The LG velocity velocity reference frame is
used for galaxies inside the C-S Cloud (using the Yahil, Tammann \& Sandage
(1977) vector for the heliocentric to LG transformation), as these galaxies
demonstrably share the LG peculiar velocities.

\section{Virgocentric Infall By the Numbers}

     The O-R-M peculiar velocity of the LG is, in the CMB frame, (-399.5,
139.3, -52.4)~km~s$^{-1}$ in (X,Y,Z) Supergalactic coordinates, or
426.3~km~s$^{-1}$ towards L = 160.8$^\circ$, B = -7.0$^\circ$ (Galactic l =
307.8$^\circ$, b = 18.2$^\circ$).  While this is essentially the same
direction as found by Lynden-Bell et al. (1988) in the original 7Samurai
analysis, the overall O-R-M velocity field predicts a lower peculiar velocity
for the LG in this direction, as well as a more complex velocity field outside
and within the LSC than the Great Attractor model used there and by
Faber \& Burstein (1988).

     In the O-R-M model, the VC has a net CMB motion relative to the LG of
-178~km~s$^{-1}$, resulting in a VC peculiar velocity {\it in the direction
of} (i.e., not relative to) the LG of $\rm Vpec(VC:LG)_{\rm O-R-M} = 90 - 178
= -88$~km~s$^{-1}$.  The O-R-M model also gives the LG a peculiar velocity
towards the direction of the VC of 227~km~s$^{-1}$.  This leaves a net
peculiar velocity of the LG {\it in the direction of} the VC of $\rm
Vpec(LG:VC)_{\rm O-R-M} = 372 - 227 = 145$~km~s$^{-1}$.

     Note that, if we insist on keeping the VC peculiar motion to be zero in
the CMB/O-R-M frame of motion, the net motion {\it between} the LG and the VC
is 145~km~s$^{-1}$.  However, if we take at face value the calculated peculiar
motion of the VC relative to the CMB/O-R-M frame (-88~km~s$^{-1}$), we get the
net motion {\it between} the LG and the VC to be $145 - (-88) = 233 \pm
63$~km~s$^{-1}$.  In other words, if we remove the outside gravitational
influences from the Virgocentric infall issue, the Local Group is moving
towards the Virgo cluster at a speed of $233 \pm 63$~km~s$^{-1}$, and the VC
itself is moving 88~km~s$^{-1}$ towards the LG.

     To model the residual velocity field within the LSC relative to the {O-R-M}
model we use a Virgocentric infall (Vinfall) model based on that used by
Lynden-Bell et al. (a linearized model of that proposed by Schechter 1980). In
this model, we nominally assign a LG Virgocentric infall velocity of
220~km~s$^{-1}$, leaving a residual of $-13 \pm 63$~km~s$^{-1}$ of LG motion
relative to the VC; i.e., well within the noise.  Likewise, the combined
O-R-M+Vinfall model leaves a net LG motion towards the center of Umaj of
+15~km~s$^{-1}$, with an error at least as large as that towards the VC.  As
one will see in the graphs at the end of the paper, the combined O-R-M +
Vinfall model accounts for all but $\sim 100$~km~s$^{-1}$ systematic motions
{\it within} the LSC, also within the known noise of galaxy peculiar motions.

\section{The Local Anomaly by the Numbers}

     The CMB motion of the LG is 592~km~s$^{-1}$ towards l = 276.0$^\circ$, b
= 23.6$^\circ$ (L = 146.0$^\circ$, B = -33.9$^\circ$). Subtracting the LG
O-R-M motion, 426~km~s$^{-1}$ towards l = 307.8$^\circ$, b = 18.2$^\circ$ (L =
160.8$^\circ$, B = -7.0$^\circ$), we get a net motion of the LG relative to
the O-R-M model of 309~km~s$^{-1}$ towards l = 228.9$^\circ$, b = 19.8$^\circ$
(L = 93.5$^\circ$, B = -63.9$^\circ$). Removing a Virgocentric infall of
220~km~s$^{-1}$ towards l = 283.8$^\circ$, b = 74.5$^\circ$ (L =
102.9$^\circ$, B = -2.4$^\circ$) yields a net motion of the LG relative to the
combined O-R-M+Vinfall models of 273~km~s$^{-1}$ towards l = 222.9$^\circ$, b
= -4.7$^\circ$ (L = 18.9$^\circ$, B = -85.3$^\circ$).

      Because both the O-R-M vector and the Vinfall vectors lie so close to
the Supergalactic plane (7$^\circ$ and 2.4$^\circ$, respectively), the
-330.1~km~s$^{-1}$ Z-component of the LG CMB motion cannot be fit by either
component.  Thus, we are left with a component of the Local Group's CMB
velocity that is larger than its Virgocentric infall velocity, in a direction
almost exactly perpendicular to the Supergalactic plane - the Local Anomaly.

      The problem with understanding this part of the Local Anomaly motion
lies, of course, with the direction it is located within our Galaxy - within
5$^\circ$ of the Galactic plane.  Given that the Supergalactic plane is nearly
perpendicular to the Galactic plane, it comes as little suprise that almost
all of the galaxies known within the LSC lie within the Supergalactic plane.

\section{The Pictorial Representations of the Model Fits}

     Here we concentrate on presenting the residual motions within the LSC
{\it after} this model is subtracted from the observed peculiar velocity
field.  The interested reader is referred to Burstein et al. (1996) for how
the peculiar velocity field appears before this model-fit. As in Burstein et
al., we restrict our picture of the peculiar of motions of LSC galaxies to
those galaxies lying within $\rm \pm22.5^\circ$ of the Supergalactic plane,
and within $\rm \pm45^\circ$ of the Supergalactic Y direction.  The planar cut
removes issues pertaining to galaxy infall relative to the Supergalactic
plane; the Y-direction cut restricts our view to those motions that are
primarily directed towards the two largest masses within the LSC --- VC and
Umaj.

    Figure 1 shows the peculiar velocity field relative to the O-R-M+Vinfall
model, as viewed from the direction of the north Supergalactic pole, with
peculiar velocities of the galaxies plotted in their full (i.e., not
{\it projected} velocities), a~la$'$ Lynden-Bell et al.  For this view alone, we
include the peculiar velocities of the galaxies known to be in the VC.

\begin{figure}
\psfig{file=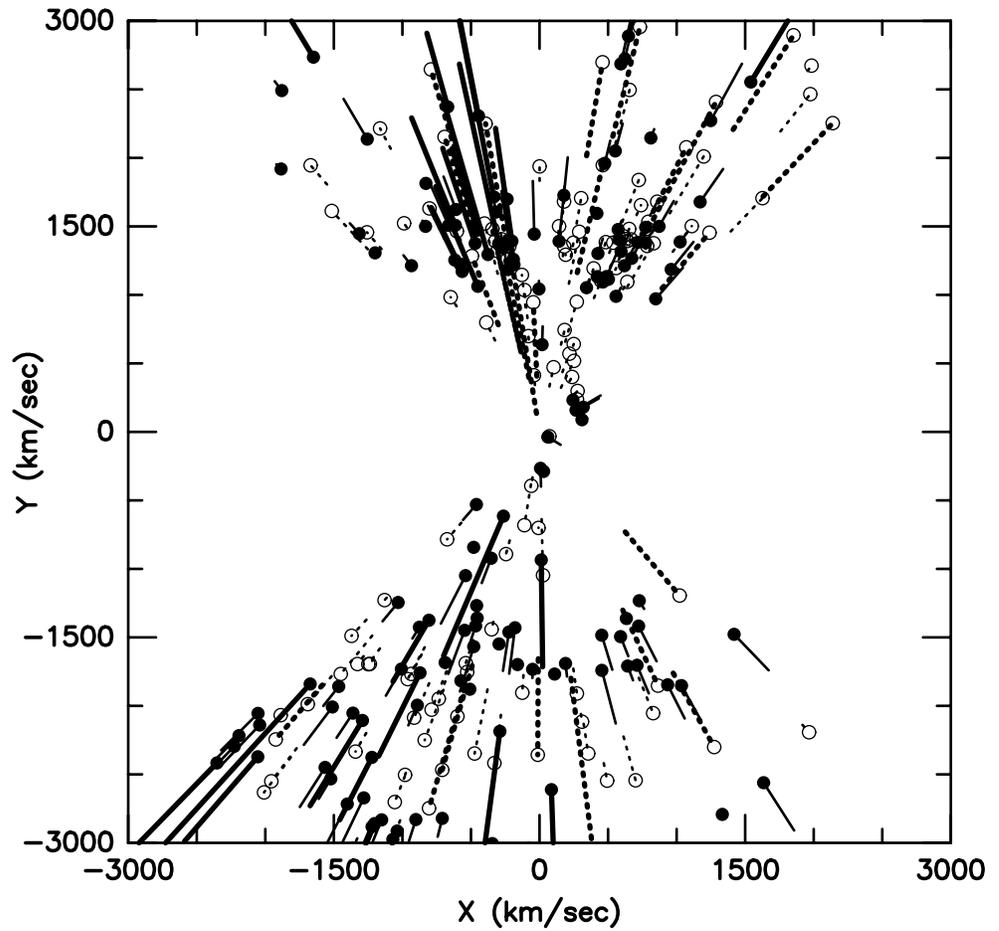}
\caption[burstein_victoriafig1.eps]{The peculiar velocity field of the Mark III
catalog for those galaxies within $\pm 22.5^\circ$ of the Supergalactic plane,
limited to $\pm 45^\circ$ of the SGY direction, relative to the combined
Outer-Region-(generated)-Model plus Virgocentric infall model as discussed in
the text.  In this model, the Virgo cluster itself is given a velocity of
88~km~s$^{-1}$ towards the Local Group. \label{fig1}}
\end{figure}

     Figure~2a shows the peculiar velocity field for the same galaxies as in
Figure~1, relative to just the O-R-M model.  Fig.~2a views the peculiar
velocities along the Y-axis as if removed from the flow, negating the sign of
peculiar velocities in the -Y direction.  Comparison of this figure with
Fig.~5a in Burstein et al. shows that the O-R-M model removes the apparent
infall towards the LG for galaxies with $\rm Y \le -1000$~km~s$^{-1}$. The
O-R-M model now makes it clear that the apparent infall velocities seen along
the Y axis are primarily a LSC phenomenon, rather than one of a larger region.

     Figure~2b shows the peculiar velocity field along the Y axis relative to
the combined O-R-M+Vinfall model velocity field (in this and subsequent views,
galaxies are excluded if they lie within 5.7$^\circ$ of the VC direction,
and within 530~km~s$^{-1}$ of the VC distance). As stated earlier, it is
apparent that the combination of the O-R-M + Vinfall velocity fields fit well
the velocity field within the LSC, modulo residual motions of $\sim
100$~km~s$^{-1}$ remaining, as evident by the coherent residual motions 
of the galaxies within the Coma-Sculptor Cloud.

\begin{figure}
\psfig{file=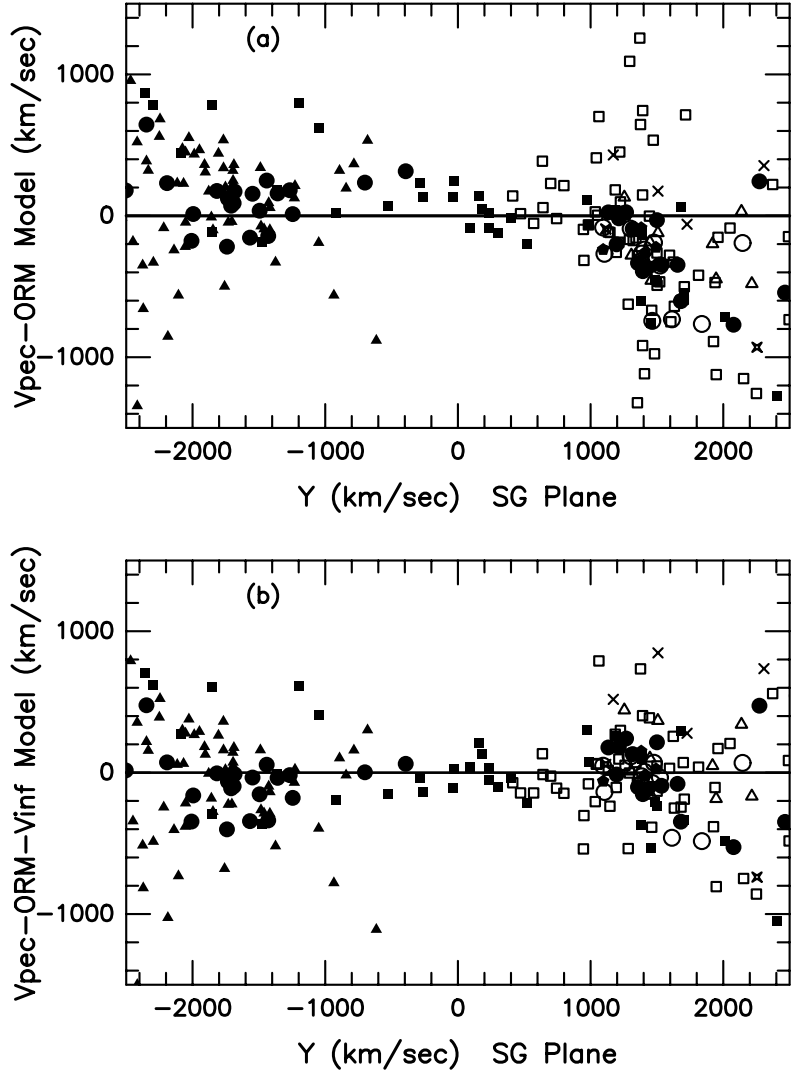}
\caption[burstein_victoriafig2.eps]{(a) The Mark III predicted peculiar
velocities relative to the Outer-Region-(geneated)-Motion model, plotted
versus SG $\pm$Y direction for the galaxies shown in Figure~1. 
The sign of peculiar velocity is negated for -Y
positions. Open symbols are galaxies within 35$^\circ$ of the Virgo cluster
direction (l = 283{\hbox{$.\!\!^\circ$}}8, b = 74{\hbox{$.\!\!^\circ$}}5);
closed symbols are galaxies in other directions.  Circles denote galaxies with
2 or more separate observations; squares denote Aaronson et al. data;
triangles Mathewson data; x's are data from Courteau and pentagons data from
Willick; see Willick et al. 1995, 1996 for details. Comparing this figure to
the analogous one in Burstein et al. 1996, one sees that the O-R-M model
removes the infall pattern for galaxies with $\rm Y < -1000$~km~s$^{-1}$
as seen in the CMB reference frame. (b) The same galaxies as
in (a), now with both the O-R-M and Virgocentric infall model subtracted.  It
is evident that the combination of the two models has removed all large-scale
flows from within the Local Supercluster. \label{fig2}} 
\end{figure}

     Figures 3a--d show the views in a 35$^\circ$ cone towards the VC
(left-hand side) and in a 25$^\circ$ cone towards Umaj (RHS).  The top graphs
are the peculiar velocities of galaxies as seen in the observed CMB + Local
Anomaly reference frame. The bottom graphs are the peculiar velocities of the
galaxies in these directions relative to the combined O-R-M+Vinfall model.  It
is clear the combined model not only removes all evidence of infall on the
near side of the VC, it also removes essentially all of the inwards motions of
galaxies in the the UMaj direction.  There is even a suggestion in Fig.~3d of
a slight infall pattern relative to Umaj itself.

\begin{figure} \psfig{file=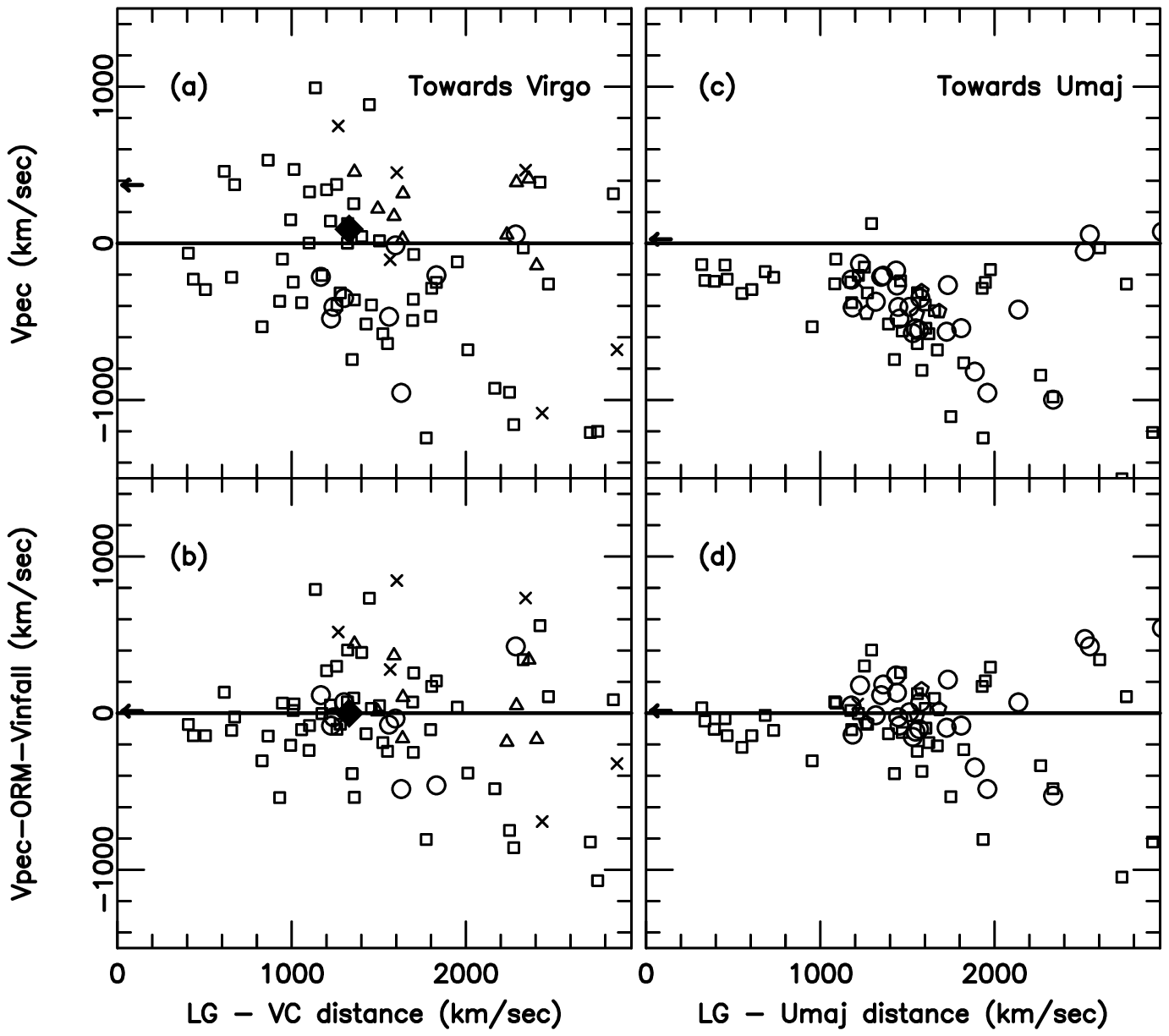}
\caption[burstein_victoriafig3.eps]{Left hand side (a,b): The peculiar motions
of galaxies from Figure~1 that are also within a 35$^\circ$ cone centered on
the direction of the VC ($\rm l = 283.8^\circ, b = 74.5^\circ$), with VC
galaxies excluded. Right hand side (c,d): The peculiar motions of galaxies
from Figure~1 that are also within a 25$^\circ$ cone centered on the direction
of Umaj (($\rm l = 140^\circ, b = 62^\circ$).  Top row (a,c): the observed CMB
+ Local Anomaly reference frame.  Bottom row (b,d): relative to the combined
O-R-M + Vinfall model.  The large infall motion of the galaxies towards UMaj
are removed by the model, as well as infall motions towards the VC. There is a
hint of an infall pattern remaining relative to UMaj.  Symbols the same as in
Fig.~2; the arrows give the peculiar velocity of the Local Group in these
directions in these reference frames. \label{fig3}}
\end{figure}

\section{Conclusion}

     The well-known quadrupole peculiar velocity field within the LSC is
well-fit by a a combination of two physically-generated motions:  a) Those
generated by mass distributed around the Local Supercluster at $D \ge
3000$~km~s$^{-1}$ (from the O-R-M model); and b) the Virgocentric infall
(Vinfall) pattern from a Virgo Cluster moving at 88~km~$^{-1}$ motion towards 
the LG, which produces 220 km~s$^{-1}$ of infall of the LG towards the VC. 
(If we insist the VC have no peculiar motion in the O-R-M model, there would be 
an LG infall velocity to the VC in the O-R-M model of 132~km~s$^{-1}$.) The 
strong observed infall motion of the Ursa Major region is well-fit by the 
combined O-R-M+Vinfall model {\it plus} the net infall motion of the Virgo 
cluster towards the LG. If one insists on putting the Virgo cluster at rest 
relative to the model, all the galaxies in the Ursa Major region are then 
given an overall 88~km~s$^{-1}$ peculiar motion towards the LG.

     Two kinds of coherent motions remain relative to the combined model: a)
$\sim 100$~km~s$^{-1}$ motions, coherent over 1000-1500~km~s$^{-1}$ size
scales (e.g., what we see for the C-S Cloud).  b) The Local Anomaly, now seen
as a 270~km~s$^{-1}$ motion of the Local Group nearly perpendicular to the
Supergalactic plane.  Is this SGZ motion just that of the Local Group, does it
involve the whole Coma-Sculptor Cloud, or does it involved the whole LSC?

While the "Local Void" (cf. Tonry \& Dressler, this meeting) is likely a
contributant to the Local Anomaly motion, for it to ``push'' the LG stronger
than the VC ``pulls'' us seems unlikely.  As the bulk of our knowledge of LSC
motions lie within the Supergalactic plane, it is improbable for us to know 
the full origin of the Local Anomaly without knowing the distances of galaxies
behind the Zone of Avoidance of the Galactic plane. Saito et al. (1991) and 
Lahav et al.
(1993) have shown that near l = 230$^\circ$, b = -4$^\circ$ is a galaxy 
cluster in
Puppis, heavily extincted by the Galaxy and masked by foreground stars.  Given
this cluster appears strongly in IRAS-detected galaxies, it is likely that it
also contains a substantial number of early-type galaxies. If we can measure
the distances to galaxies in the Puppis region, we can test whether the SGZ
motion of the LG is shared by the whole LSC or not.  If it is just the LG/C-S
Cloud doing this motion, then we should simply see the reflex motion of the SGZ
reflected in the peculiar velocities of the Puppis galaxies.  If, on the other
hand, the LSC and the Puppis Supercluster are being gravitationally drawn
together, the galaxies in Puppis will show peculiar velocities towards us in
substantial excess of the CMB reflex.  This is a test definitely worth doing,
albeit the known difficulties involved.

\acknowledgements

     This work could not have been done withont the efforts of the whole
Mark III team (Jeff Willick, Sandy Faber, Stephane Courteau, Avishai Dekel,
Michael Strauss, Tsafrir Kolatt and Bepi Tormen), and especially the O-R-M
calculations provided to the author by Ami Eldar and Avishai Dekel.  My 
heartfelt thanks to all of them.

\end{document}